\documentclass[prd,aps,twocolumn,nofootinbib,preprintnumbers]{revtex4}
\usepackage{amsmath,amssymb,epsf}
\usepackage{graphicx}

\def\be{\begin{equation}}
\def\ee{\end{equation}}
\def\bea{\begin{eqnarray}}
\def\eea{\end{eqnarray}}

\def\la{~\mbox{\raisebox{-.6ex}{$\stackrel{<}{\sim}$}}~}
\def\ga{~\mbox{\raisebox{-.6ex}{$\stackrel{>}{\sim}$}}~}

\def\ltap{\ \raise.3ex\hbox{$<$\kern-.75em\lower1ex\hbox{$\sim$}}\ }
\def\gtap{\ \raise.3ex\hbox{$>$\kern-.75em\lower1ex\hbox{$\sim$}}\ }
\def\gl{\ \raise.5ex\hbox{$>$}\kern-.8em\lower.5ex\hbox{$<$}\ }
\def\roughly#1{\raise.3ex\hbox{$#1$\kern-.75em\lower1ex\hbox{$\sim$}}}

\newcounter{oldcounter}
\addtocounter{equation}{1}
\begin{document}

\title{Confronting pNGB quintessence with data}
\author{Koushik Dutta\footnote{\tt
koushik@physics.umass.edu} and
Lorenzo Sorbo\footnote{\tt sorbo@physics.umass.edu}}
\affiliation{Department of Physics,
University of Massachusetts, Amherst, MA 01003}
\date{\today}

\begin{abstract}
We analyze the observational constraints on the model where a pseudo-Nambu-Goldstone boson (pNGB) plays the role of dark energy. The constraints are derived by using the latest Gold set of 182 type Ia supernovae and the CMB shift parameter. We allow for both the initial value of the scalar field and the present value of the energy density in the pNGB to vary. We find that -- compared to previous analyses -- the allowed portion of parameter space has shrunk around the region where the pNGB does not evolve significantly.

\end{abstract}

\preprint{{\tt astro-ph/0612457}}

\maketitle

\tighten

\draft

\section{Introduction}

Since the type Ia supernova observations of~\cite{riess98,sne,Riess}, an intense activity has been devoted to the search of an explanation of the accelerated expansion of the Universe. If gravity is described by Einstein's General Relativity and the effects of inhomogeneities can be neglected, then acceleration must be due to a dark energy component that represents roughly 70\% of the matter content of the Universe. Current data tell that the equation of state parameter $w=p/\rho$ of dark energy has to obey $w\la -0.7$~\cite{Riess:2006fw}. 

The simplest explanation of cosmic acceleration is a cosmological constant, the energy of vacuum, with magnitude 
\begin{equation}\label{lambda}
\Lambda\simeq (2\times 10^{-3}\, {\mathrm {eV}})^4
\end{equation}
and equation of state parameter $w=-1$. This solution is attractive in many respects, both for its simplicity (a single parameter is needed to describe it) and for its excellent agreement with data. It is however hard to justify from a theoretical standpoint. Quantum fluctuations of matter, indeed, give contributions to the vacuum energy, and very precise cancellations are needed to keep this energy at small values. Unfortunately the Standard Model does not display any of these cancellations at least up to the scales that have been probed in collider experiments, about $60$ orders of magnitude beyond the value of~(\ref{lambda}).

For this reason, soon after the release of~\cite{riess98,sne,Riess}, people have started to look for alternative scenarios, and a wide interest in {\em quintessence} models has emerged~\cite{q} (for a recent review, see~\cite{Copeland:2006wr}). The philosophy behind quintessence is the following. First, it is assumed that some mechanism is able to fix the energy of the ground state of the Universe to zero\footnote{It is often stated that finding a mechanism that fixes to cosmological constant to zero should be easier than finding a mechanism that fixes it to some very small nonvanishing value. Let us note here that this is not what usually occurs in Quantum Field Theory: if it is possible to find a symmetry that fixes some quantity to zero, it is typically straightforward to break such a symmetry so that this quantity can be kept small in  a controlled way.}. Then the existence of a new degree of freedom  (quintessence) is postulated: quintessence is supposed not to have yet relaxed to its vacuum, so that its energy density is responsible for cosmic acceleration. Quintessence has $w\neq -1$ as a distinctive prediction, and is usually described by some scalar degree of freedom $\phi$ endowed with some potential $V\left(\phi\right)$. $V\left(\phi\right)$ has to be very flat, if we want $w$ to be sufficiently negative.

It is possible to write down a virtually infinite number of quintessence potentials $V\left(\phi\right)$. However, only for few of them the flatness of the potential is not spoiled by radiative corrections and the exchange of quanta of $\phi$ do not give rise to an (unobserved) fifth force~\cite{quintconst}. Those few potentials are more motivated from a theoretical point of view than the others. This is especially true for the pseudo-Nambu-Goldstone boson (pNGB) potential of~\cite{fhsw} that has all the good qualities of radiative stability that anybody who believes in quantum mechanics might require.

In the present paper we analyze the parameter space of  pNGB quintessence~\cite{fhsw} in the light of the most recent observations, in particular those from supernovae. Our approach is orthogonal to the ``model independent'' approach recently taken on the subject by many investigators (see for instance~\cite{modind}), and is admittedly based on a theoretical prejudice in favor of radiatively stable potentials. To 
our knowledge, the most recent complete analyses of this model date back to about five years ago~\cite{Waga:2000ay,Ng:2000di,kmt}. Given the recent developments of the observational situation, we believe that it is important to perform an analysis of the model in which the latest data are taken into account.

In the next section we briefly describe the properties of the model of pNGB quintessence. Then in section III we present the observational constraints from supernovae and from the CMB shift parameter. In section IV we discuss our results before concluding in section V.

\section{The pNGB potential}

The use of pNBGs has been first proposed in order to realize a technically natural model of inflation in~\cite{Freese:1990rb} and has been subsequently considered for dark energy in~\cite{fhsw}. The model is characterized by a pseudoscalar field $\phi$ with a potential that can be well approximated by
\begin{equation}\label{vofphi}
V\left(\phi\right)=\mu^4\,\left[\cos\left(\phi/f\right)+1\right]\,\,,
\end{equation}
where we have neglected the contributions by higher harmonics (this is supposed to be a good approximation as long as $f$ is sufficiently smaller than the Planck mass~\cite{badifogo}). The potential is generated by the breaking of a shift symmetry $\phi\rightarrow \phi+$constant, and for this reason it is radiatively stable.

The cosmological evolution of this model is in general determined by four parameters: the quantities $\mu$ and $f$ and the initial conditions $\phi_{\mathrm {in}}$ and $\dot\phi_{\mathrm {in}}$. Due to the high expansion rate of the Early Universe, we assume $\dot\phi_{\mathrm {in}}=0$. One more free parameter is eliminated if we insist that $\Omega_\phi$ (the current ratio of the amount of dark energy over critical energy) is equal to $0.7$. As a consequence, if we assume $\Omega_\phi=0.7$, the model is described by only two parameters that can be taken to be $f$ and $\phi_{\mathrm {in}}$. A detailed analysis of the dynamics of the pNGB zero mode can be found in~\cite{Ng:2000di}. Due to periodicity of the potential $\phi_{\mathrm {in}}$ takes values between $0$ and $2\pi f$. In addition, if we take into account the indication from String Theory~\cite{badifogo} that $f$ cannot be larger than $M_P\simeq 2.4\times 10^{18}$~GeV, then the parameter space of the potential turns out to be compact. This implies that, at least in principle, all of it can be excluded by observation, and that pNGB quintessence in its simplest version can be ruled out. For this reason we find this model even more attractive (although there are ways to evade the constraint $f<M_P$~\cite{largef}), and we believe this is an additional motivation for studying it in detail. Also, for this reason we will restrict our study to the region $f<M_P$.

\section{Observational Constraints}

We consider the quintessence field $\phi$ with potential~(\ref{vofphi}) 
in a flat Friedmann-Robertson-Walker Universe. The equations of motion are given by
\begin{eqnarray}\label{bckgrnd}
&&H^2 = \frac{1}{3M^2_{P}}(\frac{1}{2}\dot \phi^2 + V(\phi) +
\rho_{m})\,\,,\nonumber\\
&&\ddot \phi + 3H\dot \phi + \frac{dV}{d\phi} = 0\,\,,
\end{eqnarray}
where $H = \dot a/a$ is the Hubble constant, $\rho_m$ is the energy density of nonrelativistic matter and, since we will be dealing only with the dynamics of the late Universe, we neglect contributions by radiation. We solve these equations numerically to find the evolution of the scale factor as a function of time.

\subsection{Type Ia supernovae} 

First, we investigate constraints from the observation of type Ia supernova from the data set~\cite{Gold182}, which
is a compilation of old data \cite{Riess} by HZS team, first year
SuperNova Legacy Survey data \cite{Astier:2005qq} and recent observations of 21 new supernovae \cite{Riess:2006fw}. For our analysis we will consider
only the 182 ``high confidence'' Gold SN data with $z> 0.0233$.
Although most of the SNe have $z <1$, there are $16$ SNe with $z>1$. This is the most up to date supernova data set available in the
literature. This data set has been recently used in~\cite{Barger:2006vc,Gong:2006gs} to study the observational constraints on different parametrizations of dark energy.

For a particular cosmological model with parameters ${\bf s}$ the
predicted distance moduli are given by
\begin{equation}
\mu_{0}(z, {\bf s}) = m - M  =
5~\log_{10}\left(\frac{d_{L}\left({\bf s}\right)}{\mathrm {Mpc}}\right) + 25,
\end{equation}
where $m$ and $M$ are the apparent and absolute magnitude of distant
supernovae. $d_{L}$ is the luminosity distance given by
\begin{equation}
d_{L}(z) = (1+z)\int_{0}^{z}\frac{dz^{'}}{H(z^{'})}
\label{luminosity_distance}
\end{equation}
and depends only on the expansion history of the
Universe from redshift $z$ to today. Assuming that all the
distance moduli are independent and normally distributed the
likelihood function can be calculated from the chi-square
statistics ${\mathcal L} \propto \exp (-\chi^2/2)$, where
\begin{equation}
\chi^2 (\phi_{in}, \dot \phi_{in}, f, \mu, H_{0}) = \sum_{i =
1}^{182} \frac{(\mu_{0i}^{obs} - \mu_{0i}^{th})^2}{\sigma_{0i}^2}.
\end{equation}
Here $\mu_{0i}^{obs}$ and $\sigma_{0i}$ are the measured value of
the distance modulus and the corresponding uncertainty for the
$i-$th supernova. $\mu_{0i}^{obs}$ and $\sigma_{0i}$, as well as the redshift $z_{i}$ are found from the data set~\cite{Gold182}.
$\mu_{0}^{th}$ is calculated by using eq.~(\ref{luminosity_distance}), 
where $H(z)$ is obtained by
numerically solving the background evolution equation~(\ref{bckgrnd}). We marginalize the likelihood over the nuisance parameter $H_0$~\cite{sne}.

\begin{figure}[h]
\centerline{
\includegraphics[width=0.5\textwidth]{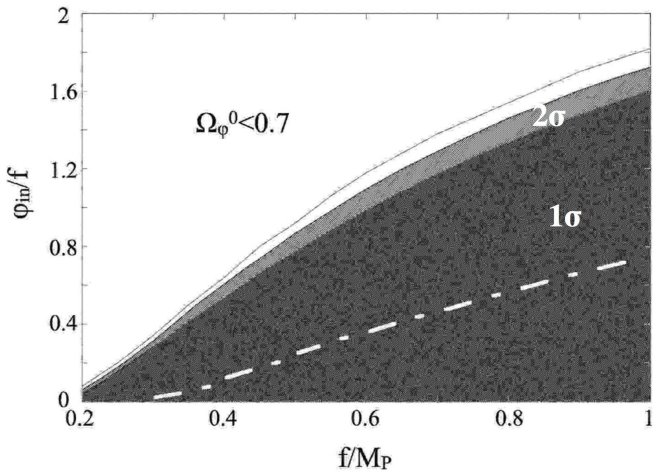}
}
\caption{The shaded areas at the bottom right of the figure denote the $1\,\sigma$ and $2\,\sigma$ confidence level regions for $\Omega_\phi=0.7$. The upper left part of the plot corresponds to parameters for which $\Omega_\phi$ never reaches $0.7$. The white dashed line corresponds to the value of the parameters for which $w_0=-0.965$.}
\label{phiinf}
\end{figure}

As we have noted earlier, the model has four parameters: $f, \mu,
\phi_{\mathrm {in}}$ and $\dot \phi_{\mathrm {in}}$. We assume $\dot \phi_{\mathrm {in}}=0$ and
allow the system to evolve until $\Omega_{\phi}
= 0.7$ today. This leaves us with two parameters and we choose
them to be $f$ and $\phi_{\mathrm {in}}$. We plot the resulting confidence
contour in figure~\ref{phiinf}. The upper left portion of the plot corresponds 
to the part of parameter space where $\Omega_{\phi}$ does not reach the value $0.7$. In this  part of the parameter space, the scalar field rolls
quickly to the minimum and oscillates around it, behaving like matter.

The dark areas at the bottom right part of the plot give the 68.3\% ($1\,\sigma$) and the 95.4\% ($2\,\sigma$) confidence level regions. The $3\,\sigma$ contour runs between the $2\,\sigma$ contour and the boundary of the forbidden region (we do not show it in the plot for clarity). As we go closer to the boundary of the forbidden region, the value of $\chi^2$ increases sharply. Even if the 95.4\% confidence level area seems to cover almost all of the allowed region, this is actually not the case. Indeed, there is a part of the parameter space, below the boundary of the forbidden region, where $\Omega_\phi$ goes across $0.7$ several times, as an effect of the oscillations of $\phi$. This means that points with the same values of $\phi_{\mathrm {in}}$, $f$ and $\Omega_\phi$ can correspond to different histories, depending on the number of times $\Omega_\phi$ has gone across the value $0.7$. We have computed the value of $\chi^2$ in the case where $\Omega_\phi$ has crossed $\Omega_\phi=0.7$ more than once and we have found that this part of the parameter space is excluded at more than the $3\,\sigma$ level.

\begin{figure}[h]
\centerline{
\includegraphics[width=0.5\textwidth]{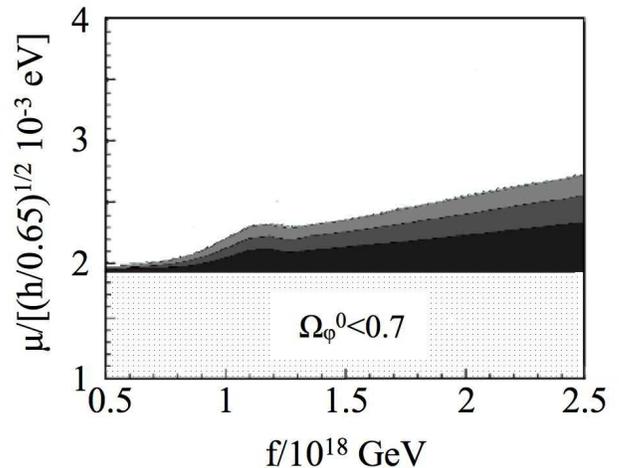}
}
\caption{$1,\,2$, and $3\,\sigma$ constraints on the plane $(f,\,\mu)$ for $\Omega_\phi=0.7$. The lower part of the plot corresponds to values of the parameters for which $\Omega_\phi$ cannot reach the value $0.7$.}
\label{figfmu}
\end{figure}

Once $\Omega_\phi=0.7$ is fixed, the constraints on the parameters $\phi_{\mathrm {in}}$ and $f$ can be converted into constraints on the plane $\left(f,\,\mu\right)$. We plot the $1,\,2,$ and $3\,\sigma$ contours in figure~\ref{figfmu}.

The thick white dashed line in the bottom right part of figure~\ref{phiinf} corresponds the part of parameter space that gives $w_0=-0.965$ (where $w_0$ is the current value of the equation of state parameter). According to~\cite{Albrecht:2006um}, $w<-0.965$ is the most optimistic constraint (at the 95.4\% level) that we might obtain from future observations, should they converge to the regime where dark energy shows no evolution. Therefore the dashed line in figure~\ref{phiinf} gives the most stringent $2\,\sigma$ constraint that we can expect to put on the parameters of pNGB quintessence.

\begin{figure}[h]
\centerline{
\includegraphics[width=0.5\textwidth]{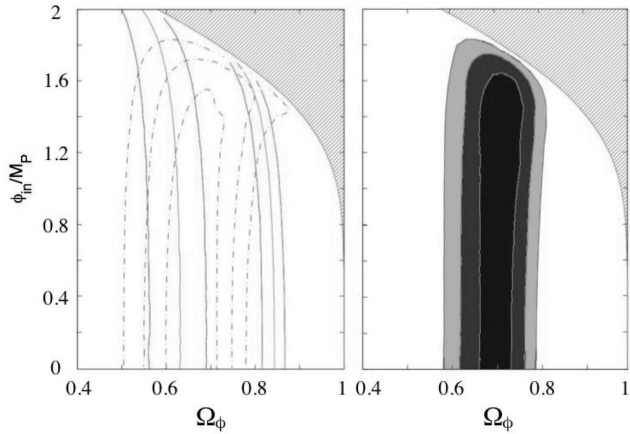}
}
\caption{Constraints on the pNGB parameter space for $f=M_P$. Left panel: $1,\,2$ and $3\,\sigma$ confidence level contours from supernovae only (dashed lines) and from the CMB shift parameter~(\ref{r}) only (solid lines). Right panel: $1,\,2$ and $3\,\sigma$ contours from the joint analysis.}
\label{omephiin}
\end{figure}

We have also considered the case where the value of $\Omega_\phi$ is allowed to vary. In this case we have fixed $f=M_P$ while keeping $\phi_{\mathrm {in}}$ variable. In the left panel of figure~\ref{omephiin} we show the $1,\,2,$ and $3\,\sigma $ contours related to supernova observations on the $(\Omega_\phi,\,\phi_{\mathrm {in}})$ plane. The shaded upper right part of the plot is excluded since the corresponding value of $\Omega_\phi$ cannot be reached. The contours are essentially vertical and centered around $\Omega_\phi\simeq 0.67$. However, at larger values of $\Omega_\phi$, somehow larger values of $\phi_{\mathrm {in}}$ (corresponding to some evolution in the quintessence field) are allowed. The best fit is at $\Omega_\phi=0.67$, $\phi_{\mathrm {in}}=0$ (so that the pNGB sits at the top of its potential and behaves as a cosmological constant) with $\chi^2=159.6$ for $180$ degrees of freedom\footnote{The best-fit $\chi^2$ for the older dataset~\cite{Riess} was of $\chi^2=178.1$ for $155$ degrees of freedom. The lower value of $\chi^2$ for the current data set can be largely attributed to more conservative assumptions on the dispersions $\sigma_{0i}$.}.
 For smaller values of $f$ (figure not shown), the contours have the same shape, although they shrink along the $\phi_{\mathrm {in}}$ direction.

\subsection{CMB shift parameter}

In addition to the SN data, we
use Cosmic Microwave Background (CMB) data to constrain the model.
In particular we derive constraints from the CMB shift parameter $\mathcal R$, that measures the shift in the angular size 
of the acoustic peaks of CMB when parameters of the theory are
varied. $\mathcal R$ is independent on the present value of the 
Hubble constant, and is given by
\begin{equation}\label{r}
\mathcal R =
\sqrt{\Omega_{m}}H_{0}\int_{0}^{z_{cmb}}\frac{dz}{H(z)},
\end{equation}
where $z_{cmb}$ is the redshift of recombination. By using WMAP 3rd 
year data the value of the shift parameter has been
extracted as $\mathcal R = 1.70 \pm 0.03$ for $z_{cmb}$ = 1089
\cite{Wang:2006ts}. In the left panel of figure~\ref{omephiin} we plot the resulting
confidence contour arising from this constraint. Note that the CMB contours favor a value of $\Omega_\phi$ that is slightly larger than that favored by the SN data.

Once the SN and the CMB constraints are combined, we obtain the plot shown on the right panel of  figure~\ref{omephiin}.  Since we are assuming a flat Universe, imposing the shift parameter constraint does not reduce significantly the area of the allowed region, and indeed the SN and CMB constraints are not orthogonal. However, the CMB constraint helps eliminate the part of parameter space at large $\Omega_\phi$ and large $\phi_{\mathrm {in}}$ that is available at the $2\sigma$ and $3\,\sigma$ level if only SN constraints are taken into account. Moreover, once the CMB constraint is added, the best fit point is not any more at $\phi_{\mathrm {in}}=0$, but at the point $\phi_{\mathrm {in}}=1.25\, M_P,\Omega_\phi=0.71$ where $\chi^2=161.6$. This should be compared to $\chi^2=162.9$ found at $\Omega_\phi=0.71$ when the constraint $\phi_{\mathrm {in}}=0$ is imposed. Since  $\phi_{\mathrm {in}}\neq 0$ implies that $\phi$ is rolling, the combination of CMB and SN data seems to hint at some evolution in dark energy. However, this hint should be taken with a grain of salt, since it emerges when we join two data sets that are not exactly compatible, as shown by the increase of $\sim 2$ units in $\chi^2$ when we add the single CMB point to the SN data set (as stated above, the best fit point for SN data only has $\chi^2=159.6$).

\section{Discussion}

The data presented above indicate that, if we use supernova constraints only, the parameters that yield the best fit to data are those where the field $\phi$ sits at the top of the cosine potential, thus mimicking a cosmological constant. If we use also the constraint from the CMB shift parameter, a slowly rolling pNGB is slightly preferred to a constant one.

\begin{figure}[h]
\centerline{
\includegraphics[width=0.4\textwidth]{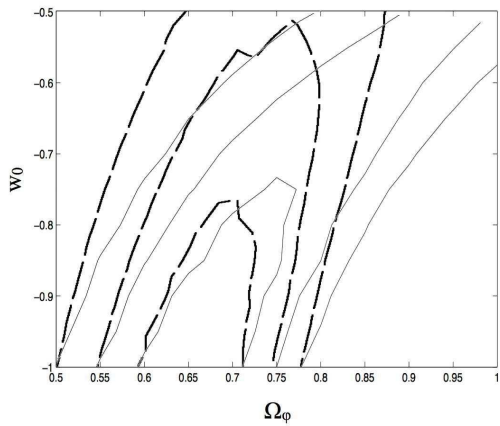}
}
\caption{$1,\,2,$ and $3\,\sigma$ contours in the plane $\left(\Omega_{\mathrm DE},\,w_0\right)$ for the pNGB model with $f=M_P$ (dashed, thicker lines) and for dark energy with constant equation of state (solid lines). }
\label{omephi}
\end{figure}

In order to see how the new data of~\cite{Riess:2006fw} improve the constraints on the model, we can compare our results with those of previous analyses. In~\cite{Waga:2000ay}, Waga and Frieman have studied the constraints from the 1998 supernova data of Riess et al.~\cite{riess98}, together with the statistics of gravitationally lensed quasars. Comparison of our results with those of~\cite{Waga:2000ay} is complicated by different assumptions on the parametrization of the model. Indeed, in~\cite{Waga:2000ay} the value of $\phi_{\mathrm {in}}$ is fixed to $1.5\,f$, so that $\Omega_\phi$ is not a free parameter, but is function of $f$ and $\mu$. Nevertheless, some comparison is possible: in the parameter space of~\cite{Waga:2000ay} there is still room at $2\,\sigma$ for a small region where the scale factor of the Universe is currently decelerating. In our analysis (see figure~\ref{omephi}) this is not possible any more at the $2\,\sigma$ level, even if it is still allowed at $3\,\sigma$.

In~\cite{Ng:2000di}, a wider portion of the parameter space is analyzed (and a  different data set~\cite{sne} is used), that shows that a part of parameter space where $\phi$ has performed half oscillation is allowed at the $2\,\sigma$ level. As we have stressed in the section III, the current data do not allow for this possibility any more.

In~\cite{kmt}, a detailed study of the parameter space of the model has been performed by taking into account the constraints from CMB observations of BOOMERanG~\cite{deBernardis:2000gy} and MAXIMA~\cite{Hanany:2000qf}. Constraints on this quintessence model were derived from its effects on integrated Sachs-Wolfe effect as well as from its effects on the location of the first peak. In figure~\ref{figfmu} we show the $1,\,2,$ and $3\,\sigma$ constraints in the $\left(f,\,\mu\right)$ plane obtained by our analysis. Comparison with figure 5 of~\cite{kmt}, shows that the more recent data improve by a factor of three or so the constraints on $\mu$.

Let us also note that a quintessence field that is climbing up the potential could mimic $w<-1$~\cite{Sahlen:2005zw,Csaki:2005vq} and possibly offer a better fit to data  (see however~\cite{Sahlen:2006dn}). In our case this is possible only if $\Omega_\phi$ has already gone through a maximum. As we have mentioned above, this case appears to be not realized at the $3\,\sigma$ confidence level for a cosine potential. Indeed, a more asymmetric potential (such as that pictured in figure 1 of~\cite{Csaki:2005vq}) is needed to impart a sufficiently large velocity to the field and improve the fit to data.

Finally, it is instructive to discuss the current value of the equation of state parameter $w_0$ as obtained in the pNGB model and to compare it with the value of $w_0$ obtained by assuming that it is constant throughout the evolution of the Universe.  In figure~\ref{omephi} we show the plots of the $1,\,2$ and $3\,\sigma$ confidence level contours in the plane $(\Omega_\phi,\,w_0)$ both for the case of pNGB quintessence with $f=M_P$ and for a model with constant $w_0$ (only supernova data are used to compute the contours in figure~\ref{omephi}). In the case of constant $w_0$ the contours are quite tilted, allowing for a value of $w_0$ significantly different from $-1$ only if $\Omega_{\mathrm {DE}}$ gets very close to unity. Indeed, if dark energy has a constant equation of state parameter different from $-1$, then a larger amount of dark energy is needed to get the same averaged value of $w$. In the case of a pNGB, however, the value of $w_0$ can be significantly different from $-1$ even if $w$ was close to $-1$ in the past. As a consequence, one can obtain the required averaged value of $w$ even without requiring that $\Omega_\phi$ is extremely close to unity. 
\section{Conclusions}
We have analyzed the portion of parameter space available for the model of pNGB quintessence. Our work extends the previous studies on the subject by allowing both for variations in the initial value of the zero mode of the pNGB and for variations in the current value of $\Omega_\phi$.
Using the most up-to-date supernova data, we have shown that the parameter space of the pNBG potential is significantly constrained around the region where quintessence is sitting on the top of the cosine potential or slowly evolving along it. At the 95.4\% level, previous analyses on the subject~\cite{Waga:2000ay,Ng:2000di} were still allowing the current value of $w$ to be larger than $-1/3$ or even the possibility that quintessence had already performed a half oscillation about its minimum. Current data do not allow this any more. 

We have also observed that, when CMB and SN constraints are joined, an evolving pNGB provides a slightly better fit to data than a pNGB stuck at the top of its potential.

Let us finally discuss future perspectives. Already now, data tell that $f$ cannot be smaller than about a third of Planck mass (unless we fine tune $\phi_{\mathrm {in}}$ to be very close to the top of the potential). As shown in figure~\ref{phiinf}, future data might constrain $f\ga M_P/2$, leading to some tension with the requirement $f\la M_P$ from String Theory~\cite{badifogo}. One might wonder if this will be enough to consider the model ``finely tuned'', and to start to consider alternative options~\cite{largef} as more natural. But one can also take a more optimistic approach: maybe future data will show that cosmic acceleration is sourced by an {\em evolving}, radiatively stable pseudo-Nambu-Goldstone boson.

\begin{acknowledgments}
We thank Subinoy Das and John Donoghue for useful discussions.  
\end{acknowledgments}

\end{document}